\documentstyle[12pt]{article}
\begin{document}
\noindent
\centerline{\large\bf NORMAL-STATE C-AXIS RESISTIVITY OF}
\centerline{\large\bf THE HIGH-T$_C$
CUPRATE SUPERCONDUCTORS}\\~\\
\centerline{N. KUMAR} \\
\centerline{Raman Research Institute, Bangalore 560080, India}\\~\\
\centerline{T. P. PAREEK and A. M. JAYANNAVAR}\\
\centerline{Institute of Physics, Bhubaneswar
751005, India}\\~\\
\baselineskip=24pt
\noindent
\centerline{{\it Abstract}}\\~\\
It is shown that a strong intraplanar incoherent scattering can
effectively block the interplanar coherent tunneling between the
weakly coupled planes of the highly anisotropic but clean
(intrinsic) materials such as the optimally doped high-T$_c$
layered cuprate superconductors. The calculated normal-state
C-axis resistivity \(\rho_c(T)\) then follows the metal-like
temperature dependence of the ab-plane resistivity
\(\rho_{ab}(T)\) at high temperatures. At low enough
temperatures, however, \(\rho_c(T)\) exhibits a non-metal like
upturn even as \(\rho_{ab}(T)\) remains metallic. Moreover, in
the metallic regime, \(\rho_c(T)\) is not limited by the maximum
metallic resistivity of Mott-Ioffe-Regel. This correlation
between the intrinsic \(\rho_c(T)\) and \(\rho_{ab}(T)\) is observed
in the normal state of the high-T$_c$ stoichiometric cuprates.\\
\vspace{1cm}

\newpage
\noindent
{\bf 1.~~Introduction.~~~}

The normal-state out-of-plane resistivity \(\rho_c(T)\) of the
high-T$_c$ layered cuprate superconductors$^1$ has raised a number
of questions,$^{2,3}$ as yet unresolved, about its temperature
dependence, its absolute magnitude, and its dependence on the
concentration of carriers, i.e., doping. Thus, while the
in-plane resistivity \(\rho_{ab}(T)\) is well known to be metallic$^1$
(i.e., with \(TCR \equiv \partial \rho_{ab}/\partial T > 0\),
and, in fact, essentially T-linear right from $T_c$ upwards to
the highest temperature of measurement) and smaller than the
Mott-Ioffe-Regel maximum metallic resistivity \(\rho_{max}^M\),
the out-of-plane resistivity \(\rho_c(T)\) shows a range of
behaviour. Thus, it has been variously reported to be
non-metallic$^{4-8}$ (\(TCR < 0)\) for underdoped samples;
mixed-metallic$^{1,9,10}$ (metallic at high temperatures but with a
non-metallic uptrun at low enough temperatures); and completely
metallic$^{5-7,9-14}$ for stoichiometric (fully oxygenated)
composition showing a T-linear 
\(\rho_c(T)\) from T$_c$ upwards, but with the absolute
magnitude of 
\(\rho_c(T) > \rho_{max}^M\) in all cases. The essential
structural feature of weakly coupled layers and the associated
large resistive anisotropy with \(\rho_c(T)/\rho_{ab}(T) \sim
10^2 - 10^5\), clearly makes the C-axis resistivity highly
sensitive to extrinsic details that presumably tend to
contaminate its intrinsic behaviour. Thus, it is entirely possible
for any measurement of the out-of-plane resistivity to pick up
some in-plane component of the resistivity tensor $\,-\,$ perhaps
externally due to misalignment of the contacts, or internally
due to the randomly distributed defects and faults providing
{\em shorts} between the weakly coupled ab-planes.$^{15}$  Such a
contamination of the out-of-plane \(\rho_c(T)\) by the in-plane
\(\rho_{ab}(T)\) can make \(\rho_c(T)\) track the metallic
temperature dependence of \(\rho_{ab}(T)\), making the former (
\(\rho_c(T)\)) an {\em apparent} metal.$^{15}$ The resistivity data on
high quality untwinned single crystals, however, strongly
suggests that the out-of-plane resistivity \(\rho_c(T)\) is
intrinsically metallic, and in fact T-linear$^{11,14}$ at least at high
temperatures, just as the in-plane \(\rho_{ab}(T)\)  is.  Its
absolute magnitude is, however, much larger. Recently, we had
proposed a mechanism that gave precisely such a behaviour. In
this mechanism$^{16}$ the inter-planar tunneling between the weakly
coupled metallic planes is  cut-off (blocked) by the
intra-planar inelastic (incoherent) scattering, leading to 
\(\rho_c(T) \propto \rho_{ab}(T)\).  This physical mechanism has
since gained a fair degree of acceptance among the workers in
the field,$^2$ while some earlier theories, linked closely to the
exotic mechanisms for high-T$_c$ superconductivity in the
strongly correlated CuO$_2$ sheets, have been argued out to be
inconsistent with known experimental facts.$^3$ Motivated by these
developments, we have re-examined the mechanism proposed by us
earlier based on simple physical arguments. In doing so we have
derived an expression for \(\rho_c(T)\) following the
Kubo-Matsubara conductivity formalism applied to a model
Hamiltonian incorporating weak interplanar tunneling and strong
intraplanar incoherent scattering. Our \(\rho_c(T)\) so derived indeed
shows a metallic behaviour with \(\rho_c(T) \propto \rho_{ab}(T)\) in
the high temperature limit, thus validating our mechanism
proposed earlier. At low enough temperatures, however, we get an
additional feature of a resistivity upturn
(\(\partial\rho_c/\partial T < 0\)), which is qualitatively consistent
with  observations as noted above.  In the following, we give
some details of our derivation, and discuss our results for
\(\rho_c(T)\)\ in the light of some recent findings.

\noindent
{\bf 2. Theoretical.~~~}

First note that the C-axis transport, except possibly for the
overdoped cuprates, is known to be incoherent,
e.g., \(\omega_{pc} \tau_{ab} \ll 1\), where \(\omega_{pc}\) is
the C-axis plasma frequency and \(1/\tau_{ab}\) is the
intraplanar inelastic scattering rate.$^{2,17}$ Thus, the successive
interplanar tunneling amplitudes are phase-uncorrelated. It is,
therefore, sufficient to consider simply a bilayer (AB) coupled
weakly by a tunneling matrix element (\(-t_c\)). In real systems
the individual layer (A or B) can by itself represent a single
CuO$_2$ sheet as in LSCO, or also a group of strongly coupled
CuO$_2$ sheets, as in YBCO, BSCCO and other multilayered
cuprates, separated by the spacer oxide layers. Also, we will
consider only the clean limit as suggested by the smallness of
the zero-temperature intercept$^{11}$ \(\rho_{ab}(T \rightarrow 0)\), and also
assume that the inter-planar tunneling conserves the wavevector
parallel to the ab-plane. Then the model Hamiltonian (in obvious
notation) is
\begin{equation}
H = H_a + H_b + H_{ab} + H_{aa} + H_{bb}\,\,,
\end{equation}
with
\begin{eqnarray*}
H_a &=& \sum_{{\bf k}\sigma} \epsilon_{\bf k} a_{{\bf k}\sigma}^\dagger
a_{{\bf k}\sigma},
\\
H_b &=& \sum_{{\bf k}c} \epsilon_{\bf k} b_{{\bf
k}\sigma}^\dagger b_{{\bf k}\sigma}\\ 
H_{ab} &=& - t_c \sum_{{\bf k}\sigma} (a_{{\bf k}\sigma}^\dagger
b_{{\bf k}\sigma} + b_{{\bf k}\sigma}^\dagger a_{{\bf k}\sigma})\\
\end{eqnarray*}
In the following we will drop the spin index $\sigma$.  
Here $H_{aa}$ and $H_{bb}$ represent the inelastic intra-planar
electron- electron scattering characteristic of the strongly
correlated two-dimensional CuO$_2$ sheets. In the present
analysis, however, these terms shall enter only implicitly and
summarily through the imaginary part of the associated retarded
electron self-energy chosen so as to be consistent with the
known T-linear in-plane resistivity \(\rho_{ab}(T)\).$^{18}$ Our problem
then is to calculate the out-of-plane resistivity \(\rho_c(T)\),
given the above \(\rho_{ab}(T)\) as an input. It is really this
connection between \(\rho_c(T)\) and \(\rho_{ab}(T)\) that
is being addressed here.

Now, for the inter-planar current operator, we have 
\begin{equation}
j_c = -ie t_c \sum_{\bf k}\,(a_{\bf k}^\dagger b_{\bf k} -
b_{\bf k}^\dagger a_{\bf k})\, .
\end{equation}
The Kubo conductivity in the dc limit is then given by (\(\hbar
= 1\))$^{19}$
\begin{equation}
\sigma = - (\frac{c}{L^2}) {\stackrel{\rm lim}{\omega
\rightarrow 0}} \frac{Im \Pi_{ret}(\omega)}{\omega}\,\,, 
\end{equation}

where $c$ = bilayer separation (the C-axis lattice constant),
$L^2$  = area of the layer, and the retarded correlation
\begin{eqnarray}
\Pi_{ret}(\omega) &=& {\rm lim} \Pi (i\omega_\nu) \\ \nonumber
&&i\omega_\nu \rightarrow \omega + i\delta\,\,\,\, (\rm analytic
\,\,\,continuation) 
\end{eqnarray}
with \(\Pi(i\omega_\nu)\), the current-current
correlation given by 
\begin{equation}
\Pi(i\omega_\nu) = 2e^2 t_c^2  \sum_{\bf k} \frac{1}{\beta}
G_a ({\bf k},\, i\omega_n) G_b({\bf k}, i\omega_n + i\omega_\nu)
\end{equation}
with
\begin{eqnarray*}
\omega_\nu &=& \frac{2\pi\nu}{\beta}, {\rm
the\,\,Bosonic\,\,\,Matsubara \,\,\,frequency}\,,\\
\omega_n &=& {\frac{(2n+1)\pi}{\beta}}\,, {\rm
the\,\,\,Fermionic\,\,\,Matsubara\,\,\, frequency}.
\end{eqnarray*} 
Here $G_a$ and $G_b$ are the temperature Green functions
for the layers A and B, respectively, in the presence of
inter-planar tunneling. For identical layers, as in
the present case, we have \(G_a = G_b = G\), say.

In the high temperature limit, i.e., for \(\hbar/\tau_{ab} \gg
\mid t_c \mid\), we can evaluate $G$ in the presence of
interplanar tunneling in terms of $G_o$, the corresponding
temperature Green function in the absence of tunneling, from the
Dyson equation
\begin{equation}
G_a\,\,=\,\,\,\,G_{oa} + G_{oa}\,\, t_c\,\, G_b t_c\,\, G_{oa}\,,
\end{equation}
giving
\begin{equation}
G_a = G_b \equiv  G = \frac{G_o}{1 - t_c^2\,G_o^2}\,\,.
\end{equation}
In writing Eqn. (6) we have used the approximation that for
\(\hbar/\tau_{ab} \gg \mid t_c \mid\) one can neglect the vertex
correction arising from the interplanar tunneling $t_c$.

Now, the intra-planar thermal Green function \(G_o({\bf k}, i\omega_n)\)
corresponds to an isolated layer (i.e., with $t_c$ = 0) and has
the general form
\begin{equation}
G_o({\bf k}, i\omega_n) = \frac{1}{i\omega_n - \epsilon_{\bf k}
- \sum({\bf k}, i\omega_n)}. 
\end{equation}
Substituting from Eqns. (8) and (7) into Eq. (5), we get
\begin{equation}
\Pi(i\omega_\nu) = 2e^2 t_c^2
\sum_{\stackrel{{\bf k},n}{\eta,\xi=\pm}} G_{o\eta} ({\bf k},i\omega_n)
G_{o\xi} ({\bf k}, i\omega_n + i\omega_\nu)
\end{equation}
Here $\pm$ refers to $G_o$  in Eq. (8) with $\epsilon_{\bf k}$
replaced by \(\epsilon_{\bf k} \pm \mid t_c \mid\).

Now we impose our condition of T-linearity of \(\rho_{ab}(T)\)
as input at the level of \(G_o({\bf k},i\omega_n)\) namely, that the
self-energy of the corresponding retarded Green function
must have an imaginary part \(\Delta(T)\) (at
the Fermi level) \(=  -Im \sum_{ret} \propto T\).  With
this input Eq. (9) together with Eq. (3) gives, after the
usual frequency summation and analytic continuation, a simple
expression for the dc conductivity
\begin{equation}
\sigma = \frac{1}{2}\, (\frac{e^2}{\hbar c})\, (c^2\nu)\,\,
\frac{\mid t_c \mid^2}{\Delta(T)}\,\, (1 + \frac{1}{1 +
(t_c/\Delta(T))^2})\,. 
\end{equation}
Here we have introduced the two-dimensional density of states
$\nu$, assumed constant. Also, $\hbar$ has been re-instated. 
This is our main result.
>From Eq. (10) it is readily seen that in the high-temperature
limit, \(\Delta(T) \gg \mid t_c \mid,\,\,\,\, \sigma \propto
1/(\Delta(T)\), or equivalently, the C-axis resistivity
\(\rho_c(T) \propto \Delta(T) \propto T\), confirming the
T-linearity of $\rho_c(T)$ at high temperatures.  It is also
clear that this mechanism giving incoherent transport along the
c-axis does not involve the usual `$k_F\ell$'-parameter characteristic
of metallic transport. Hence, \(\rho_c(T)\) is not subject to
\(\rho_{max}^M\). 

Next, we consider the low-temperature limit, \(\hbar/\tau_c <
\mid t_c \mid\). Now, we must diagonalize the tunneling
Hamiltonian \(H_t \equiv H_a + H_b + H_{ab}\) first, and
then use the T-linearity of \(\rho_{ab}(T)\) as an input at the
level of the layer-diagonal Green function. We have
\[
H_t = \sum_{\bf k} (\epsilon_{\bf k} - \mid t_c \mid)
\alpha_{\bf k}^\dagger\,\, \alpha_{\bf k} +
\sum_{\bf k} (\epsilon_{\bf k} + \mid t_c \mid) \beta_{\bf
k}^\dagger \,\, \beta_{\bf k}\,\,,
\]
where \(\alpha_{\bf k} = (a_{\bf k} + b_{\bf k})/\sqrt{2},
\,\,\beta_{\bf k} = (a_{\bf k} - b_{\bf k})/\sqrt{2}\).
The inter-planar current operator
\[
\hat{j} = ie t_c \sum_{\bf k}\, (\alpha_{\bf k}^\dagger\,
\beta_{\bf k} - \beta_{\bf k}^\dagger\, \alpha_{\bf k}).
\]
Repeating the earlier steps with this current operator, we
now get
\[
\sigma = \frac{1}{2} (\frac{e^2}{\hbar c}) (c^2\nu)
\frac{\Delta(T) t_c^2}{t_c^2 + \Delta^2(T)}
\]
Thus, at low enough temperatures we get an upturn for $\rho_c$
because $\Delta(T) \propto T$.  This upturn has been noticed as
discussed earlier. It must be emphasized here that this upturn
is a consequence of our assumption of conservation of wavevector
parallel to the planes in the tunneling process. This leads to a
hybridization gap that 
suppresses the overlap of the spectral functions corresponding
to $G_{o+}$ and $G_{o-}$.

\newpage
\noindent
{\bf 3. Discussion.~~~}

We would now like to comment on a recent
attempt to explain 
away the metallicitiy of $\rho_c(T)$ as a thermal expansion
effect.$^{20}$ Based on a recent study of the pressure dependence of
\(\rho_c(T)\) and the lattice constant `c' in \(La_{2-x} Sr_x
CuO_4\), it has been argued that the observed metallicity
(i.e., TCR $>$ 0) merely reflects a thermal expansion effect, i.e., 
\[
TCR \equiv (\partial\rho_c/\partial T)_P = (\partial
\rho_c/\partial P)_T\,\,(\partial c/\partial P)_T^{-1}\,\,
(\partial c/\partial T)_P\,, 
\]
and, therefore, the observed metallicity is only {\em apparent}.
While some such expansion effect cannot be ruled out, we find
the argument somewhat flawed for the following reason.  The
thermal expansion $<\delta c >$ of the lattice parameter `c', or
more precisely $< \delta B >$ of the width `B' of the potential
barrier for the rate-determining tunneling, is a 
mean-anharmonic effect that, of course, tends to diminish the
tunneling rate. There is, however, also a thermal fluctuation about
this mean-value which is of comparable magnitude, if not larger.
Now, inasmuch as the tunneling time is expected to be much
shorter that the typical lattice vibrational time period, we must
average the instantaneous tunneling rate over the anharmonic
fluctuations of $\delta B$. As the tunneling rate depends
exponentially on $\delta B$, overall it is expected to produce an
enhancement of the tunneling rate with increasing temperature,
and hence a negative, rather than positive TCR. 

Finally, a remark on the recent attempts$^{21,22}$ to explain the
resistivity upturn in terms of pre-existing real-space pairs
(Bosonic) as precursor to their condensation at $T_c$.  Such a
system of charged Bosons in the normal state is, however, expected to
exhibit large, universal diamagnetism which is not reported.

In conclusion, the present  microscopic-treatment of the C-axis
resistivity supports the mechanism proposed by us earlier, namely
that the strong intra-planar incoherent scattering cuts-off the
interplanar tunneling, and thus correlates \(\rho_c(T)\) with
\(\rho_{ab}(T)\).

\end{document}